\documentclass[preprint,12pt]{elsarticle}
\usepackage{amssymb}
\usepackage{amsmath}
\usepackage[colorlinks,linkcolor=cyan,citecolor=cyan]{hyperref}
\usepackage[ruled,linesnumbered]{algorithm2e}

\begin{document}
\begin{frontmatter}
\title{Adaptive high-precision sound source localization at low frequencies based on convolutional neural network}
\author{Wenbo Ma}
\author{Yan Lu\corref{cor1}}
\ead{luy@sustech.edu.cn}
\author{Yijun Liu\corref{cor1}}
\ead{liuyj3@sustech.edu.cn}
\cortext[cor1]{corresponding authors}
\affiliation{organization={Department of Mechanics and Aerospace Engineering, Southern University of Science and Technology},
	city={Shenzhen},
	postcode={518055}, 
	state={Guangdong},
	country={China}}
\begin{abstract}
Sound source localization (SSL) technology plays a crucial role in various application areas such as fault diagnosis, speech separation, and vibration noise reduction. Although beamforming algorithms are widely used in SSL, their resolution at low frequencies is limited. In recent years, deep learning-based SSL methods have significantly improved their accuracy by employing large microphone arrays and training case specific neural networks, however, this could lead to narrow applicability. To address these issues, this paper proposes a convolutional neural network-based method for high-precision SSL, which is adaptive in the lower frequency range under 1kHz with varying numbers of sound sources and microphone array-to-scanning grid distances. It takes the pressure distribution on a relatively small microphone array as input to the neural network, and employs customized training labels and loss function to train the model. Prediction accuracy, adaptability and robustness of the trained model under certain signal-to-noise ratio (SNR) are evaluated using randomly generated test datasets, and compared with classical beamforming algorithms, CLEAN-SC and DAMAS. Results of both planar and spatial sound source distributions show that the proposed neural network model significantly improves low-frequency localization accuracy, demonstrating its effectiveness and potential in SSL. 
\end{abstract}
\end{frontmatter}

\section{Introduction}

The ability to localize sound sources is a crucial part of signal processing, which aims to understand the energy and information contents that sound carries. For example, it is revealed by SSL that the major noise-generation mechanism in wind turbines is the aerodynamic interaction between the turbulence boundary layer and the trailing edge \cite{lee2012case}. It is also shown that SSL can be used to construct noise source maps which help identify not only the moving source such as the pass-by of a car but also the strength of the source \cite{ballesteros2015noise}. Another important application of SSL is fault detection in rotating machine, which is usually conducted through vibration analysis, however, it can be complemented by SSL based acoustic imaging technique with contactless measurements and added spatial dimensions \cite{cabada2015acoustic}. Recent advances in natural language processing and artificial intelligence seek to extract information embedded in human voices, such as acoustic traits and meaning of words, could also benefit from SSL due to the additional spatial information, which appears as the direction of arrival of sound sources in these cases. Relevant applications ranges from speech separation \cite{chazan2019multi,songgong2020robust} to human-robot interaction \cite{liu2018sound,li2016reverberant}. The aforementioned wind turbine and moving car noises show dominant noise in the lower frequency range ($\leq$ 1kHz), and it happens to partially overlap the lower sampling range of human speech voice in the study of reverberant sound localization with a robotic head \cite{li2016reverberant}. These research in low frequency SSL could further facilitate the development of effective noise control strategies, operations monitoring methods, and intelligent robots.

SSL methods are based essentially on microphone array signals and associated algorithms. While microphone array design is a topic worth in-depth investigation for optimized reception, here we focus on the development of localization algorithms. The performance of SSL algorithms are characterized by several parameters, including resolution, maximum sidelobe levels and spatial aliasing, and it is also affected by the presence of multiple sources and background noise in received signals and pseudo-sources from reflections in a reverberant environment \cite{merino2019review,chiariotti2019acoustic}. Continuing efforts have been put to understand and improve these affecting parameters. Conventional beamforming (CB) method is based on delay-and-sum (DAS) algorithm and it provides a visually intuitive representation of the sound strength distribution in the direction of sound sources \cite{chiariotti2019acoustic,yardibi2010uncertainty}. Although it is one of the widely used SSL methods, its quality suffers from low resolution at low frequencies due to Rayleigh resolution limit \cite{merino2019review,rayleigh1879xxxi}, and spurious sidelobes at high frequencies due to low spatial sampling rate of microphone array compared to the wavelength \cite{merino2019review}. The formulation of DAS can be viewed as a convolution between the true source distribution and a point spread function determined by the array and the scanning grid \cite{herold2015approach}, and therefore, iterative numerical methods such as DAMAS \cite{brooks2006deconvolution} and CLEAN-SC \cite{sijtsma2007clean} have been devised to extract the unknown source locations via deconvolution, and consequently, the performance of CB is improved by the removal of sidelobes and the enhancement of acoustic imaging resolution. However, DAMAS uses Gauss-Seidel method for the iterations, which require considerable computational resources, and therefore, not ideal in situations where high imaging resolution is needed. In reverbrant environment, average beamforming method is proposed for internal noise source mapping in aircraft fuselage and helicopter cockpit \cite{castellini2013average}, where the effect of pseudo-source is cancelled by moving the microphone array around and averaging over the received signals. In situations where clear source separation is plagued by background noise, a combination of target beamforming method and stochastic maximum likelihood method is developed for hearing aids enhancement \cite{zohourian2017binaural}, which exploits the adaptability of statistical model based methods under noisy conditions. 

Recently, deep learning (DL) has gained extensive interest across diverse domains. Functioning as a data-driven approach, DL has the notable ability to approximate intricate relationships between input features and output labels. For DL-based SSL, the input could be amplitude-, phase-, and covariance matrix-based which contains amplitude, phase and interchannel information, however, it appears that lower level amplitude- or phased-based features performs better than covariance matrix-based features \cite{krause2021feature}. There is a latest trend to feed raw data to the deep neural networks (DNN) and build end-to-end applications, which depends on DNN's ability to discover deep patterns among these data and hence requires less prior knowledge on signal processing \cite{vecchiotti2019end,pujol2021beamlearning}. Depends on the type of output labels, DNN can be applied to solve for two kinds of problems, classification and regression. In classification problems, the SSL output label is the index or coordinate of a subregion in the scanning grid where the source is located \cite{chazan2019multi,ma2019phased}. Whereas in the regression problem, the output is directly given as the cartesian coordinates \cite{shimada2021accdoa}, or azimuthal and elevation angles \cite{lee2022deep}. While a significant number of study has based on free field assumptions, Grumiaux et al. \cite{grumiaux2022survey} have done a comprehensive survey on DL-based SSL in indoor environment. It summarizes technical aspects including model architecture, data acquisition and learning strategies, and provides a fresh perspective for addressing challenges in acoustic beamforming-based SSL. Some notable applications of DL include localization with multi-frequency signals in the uncertain ocean environments using a single hydrophone \cite{niu2019deep},  and convolutional neural networks (CNN) based SSL using phased microphone array data at high frequency range \cite{ma2019phased}.

While DNN has shown good performance which encourages research in DL-based SSL, the localization ability of neural network model at low frequencies remains undetermined. Xu et al. \cite{xu2021acoustic} addresses the challenges of SSL at frequencies above 800Hz and reveals a superior localization accuracy compared to conventional algorithms. However, the investigation involves training various neural network models at different frequencies while maintaining a fixed distance between the sound source and the microphone array. In practical scenarios, however, the frequency of the sound source and the distance between the scanning grid and microphone array may vary. In such cases, the selection of appropriate trained model becomes problematic, hence compromising the SSL process. 

\begin{figure}[htp]
    \centering
    \includegraphics[width=0.9\linewidth]{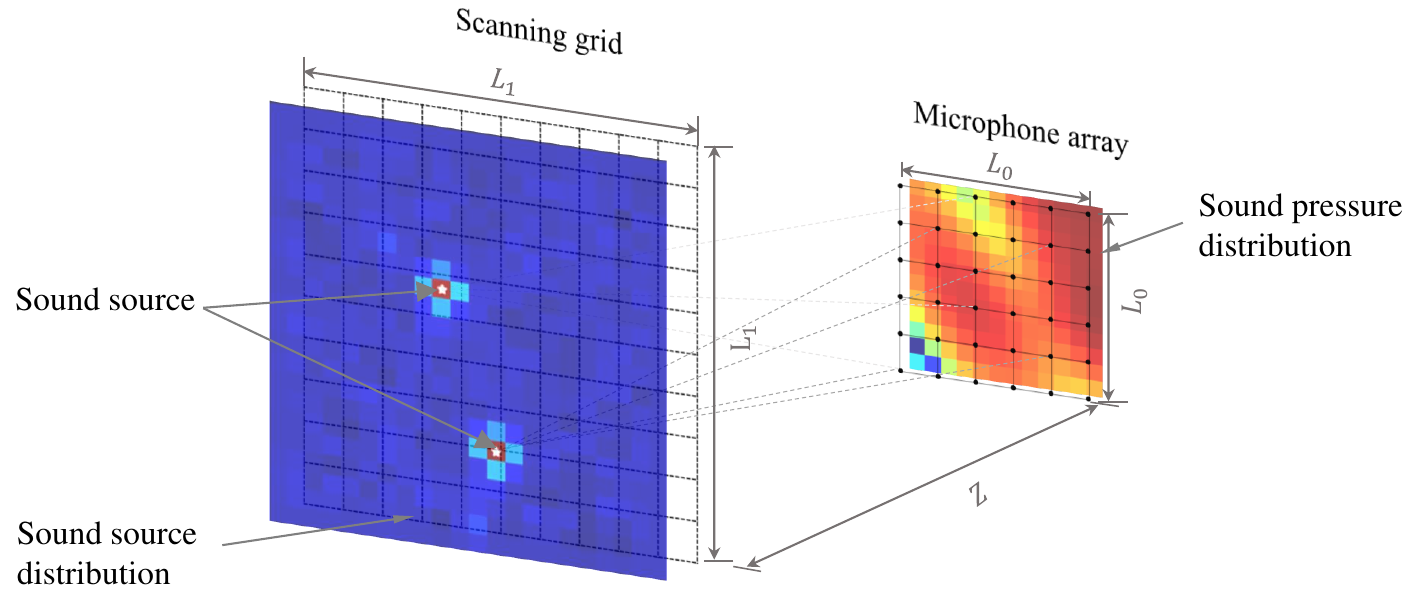}
    \caption{Schematic of the SSL problem.}
    \label{fig:planeSimulationStructure}
\end{figure}

This study employs a CNN-based model \cite{bengio2017deep} to significantly reduce trainable parameters such that construction of a deep nets is made possible. This is appealing to problems like SSL, where the model undergoes training using a substantial volume of simulated acoustic source data. The SSL problem is illustrated in Figure~\ref{fig:planeSimulationStructure}, where the sound field generated by two sources projects a partial interference image in the area of microphone array. One of the conventional measures to enhance SSL resolution would be to increase the array size, which helps receiving more sound field information.  The proposed CNN model utilizes the signals received from a relatively small microphone array arranged on a square grid with side length $L_0=0.3 m$ to predict the sound source distribution on a square scanning grid with side length $L_1=1.0 m$ at a variable distance, $Z$. This results in a significant improvement on the resolution limit in the range of 200Hz$\sim$1000Hz. For the purpose of tuning the adaptability of the network, we integrate datasets that encompass a range of conditions, including variations in sound source numbers, frequencies, and distances. Furthermore, the training labels reflecting both the location and number of sources are constructed within the deep learning framework, and a specially customized loss function is employed to improve SSL accuracy. This enhances the network's proficiency in addressing a wide spectrum of scenarios, as elaborated in the subsequent sections.

\section{Methodology}
This section discusses the input and output features, the architecture of CNN-based neural network model, the formulation of loss function, as well as source location extraction and accuracy evaluation algorithms.

\subsection{Sound field simulation}

The sound field generated by complex sound sources can be equivalently represented as the superposition of sound fields generated by multiple monopole sources in space \cite{lin2021acoustic}. Consequently, they are utilized as the sound sources in this investigation. The sound pressure on a microphone $m$ generated by a monopole source $n$ is represented as $p_n(m)$ which can be calculated using
\begin{equation}
	p_n(m)=\frac{Q e^{-j2 \pi f r/c_0}}{4\pi r},
	\label{eq:soundPressure}
\end{equation}
where $Q$ is the strength of sound source which equal to 1 in this study, $c_0$ is the speed of sound in air ($343 m/s$), $f$ is the frequency of sound source and $r$ is the distance from the microphone to sound source. In a field of multiple sources, the sound pressure received by a microphone array is calculated via the sum of pressure contributed by each source. 
\begin{equation}
	\mathbf{P} = \left\{{\sum_{n=1}^{N} p_n(1), \sum_{n=1}^{N} p_n(2), \ldots, \sum_{n=1}^{N} p_n(M)}\right\},
	\label{eq:micPressure}
\end{equation}
where $M$ denotes the total number of microphone array. The sound field expressed in Eq.~\eqref{eq:soundPressure} is in the form of complex values, however, they must be converted into real numbers for the neural network, and this is achieved by calculating the Sound Pressure Level (SPL) from collected data. SPL is commonly used in real-world scenarios and it is defined in terms of the complex norm of acoustic pressure as 
\begin{equation}
    p_m=20\log_{10}\left(\frac{|\sum_{n=1}^{N} p_n(m)|}{p_{ref}}\right),
    \label{eq:SPL}
\end{equation}
where $p_m$ denotes the SPL value of microphone $m$ in the microphone array, $p_{ref}$ denotes the reference sound pressure, usually $20\mu Pa$ in air. These SPL values are then used as the input of the subsequent neural network.

\subsection{Convolutional neural network}

A general architecture of CNN consists of an input layer, several convolutional layers, pooling layers, activation layers, fully connected layers, and an output layer. In this study, the convolutional layers are configured to employ a $3\times3$ convolution kernels, with padding to maintain the size of the data during the convolution process. Following each of the first two convolutional layers, a $2\times2$ maximum pooling step is implemented to reduce the number of features. A parametric rectified linear unit(PReLU) function \cite{he2015delving} is used after each convolutional layer to enhance the network's capability of fitting nonlinear functions. In order to regularize the network and enhance its generalization ability, batch normalization(BatchNorm) layers \cite{ioffe2015batch} are incorporated into the first three convolutional layers. A high learning rate is allowed for and it leads to accelerated convergence. Finally, fully connected layers project the features extracted from the convolutional layers to the output space. The specific structural parameters of the CNN utilized in this study are detailed in Table~\ref{tab:CNN}.
\begin{table}[!ht]
    \centering
    \caption{Parameters and structures of the CNN model}
    \resizebox{\textwidth}{21mm}{
    \begin{tabular}{lllllllll}
	\hline
        Type & Filter Size & Stride & Padding & Output Size & Activation & Normalization \\ 
        \hline
        Convolutional & $3\times3$ & 1 & 1 & $t\times t\times 64$ & PReLU & BatchNorm  \\ 
        Max Pooling & $2\times2$ & 1 & - & $\frac{t}{2}\times \frac{t}{2}\times64$ & - & -  \\ 
        Convolutional & $3\times3$ & 1 & 1 & $\frac{t}{2}\times \frac{t}{2}\times128$ & PReLU & BatchNorm  \\ 
        Max Pooling & $2\times2$ & 1 & - & $\frac{t}{4}\times\frac{t}{4}\times128$ & - & -  \\ 
        Convolutional & $3\times3$ & 1 & 1 & $\frac{t}{4}\times\frac{t}{4}\times256$ & PReLU & BatchNorm  \\
        Convolutional & $3\times3$ & 1 & 1 & $\frac{t}{4}\times\frac{t}{4}\times512$ & PReLU & -  \\ 
        Fully Connected & - & - & - & - & PReLU & - &  \\ 
        Fully Connected & - & - & - & - & PReLU & - &  \\ 
        Fully Connected & - & - & - & - & - & -  \\ 
        \hline
    \end{tabular}}
    \label{tab:CNN}
\end{table}

\begin{figure}[h]
    \centering
    \includegraphics[width=1.0\linewidth]{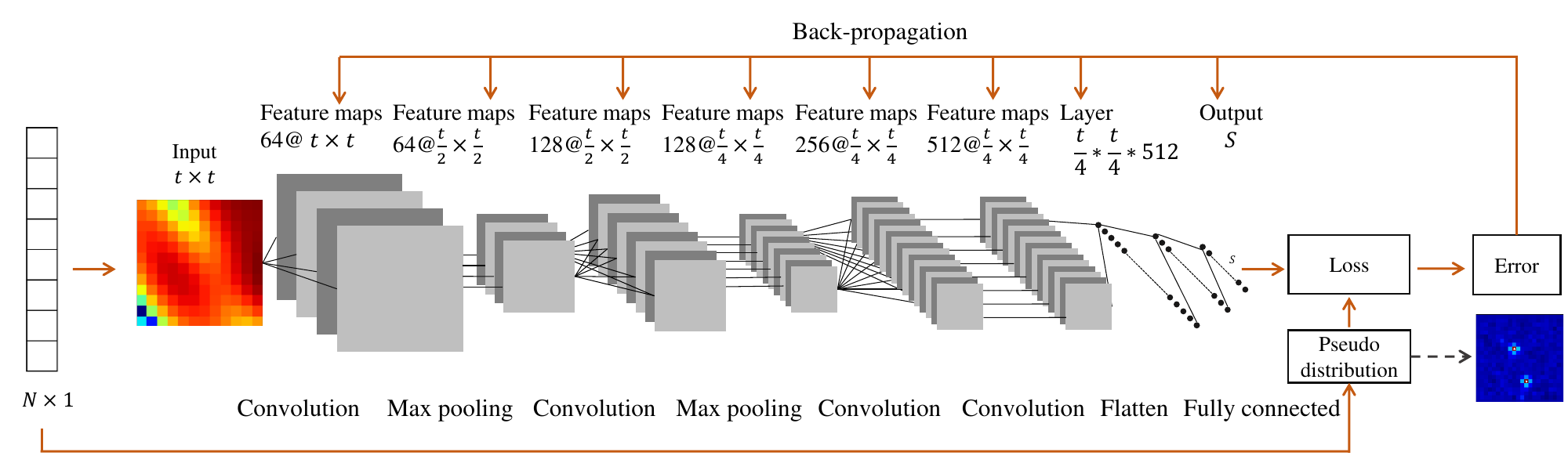}
    \caption{Schematic of the proposed CNN model.}
    \label{fig:trainingProcess}
\end{figure}

The training process is illustrated in Figure~\ref{fig:trainingProcess}. The model input is the SPL values captured from $N\times 1$ randomly distributed sound sources using a microphone array. These values are then transformed into a $t\times t$ input matrix, with $t=12$ representing the number of microphones along each array axis in this study. It can be seen that the input matrix forms a fairly complicated image, from which it is not intuitive to determine the source location at the scanning grid area. The distribution of sound sources on the scanning grid, denoted as $S$, is formulated as a $s\times s$ matrix ($s=22$), serving as the output of the neural network. Loss function calculates the error between the prediction and the true label, the specific form of which will be detailed in the subsequent section. In the optimization process, the Adam optimizer \cite{kingma2014adam} is employed to minimize the loss function. Moreover, the training dataset comprises approximately 200K samples, randomly generated to encompass various sound source numbers, frequencies, and distances between the microphone array and scanning grid, thereby enhancing the neural network's adaptability. Training procedure is executed on a cluster equipped with an Intel Xeon Gold 6148 CPU with 384GB memory and Nvidia V100 GPU.

\subsubsection{Input feature}

To exploit the inherent feature extraction capability of the CNN, we use SPL data directly converted from microphone pressure data as the input for the network. However, substantial variability in both frequency and the number of sources impedes the convergence of the neural network. To tackle this challenge, data preprocessing is incorporated to put the input data into a uniform format, which helps the network understand the data and improves efficiency. The first step of preprocessing involves Z-Score standardization, which is defined as
\begin{equation}
    x^{\prime}=\frac{x-\mu}{\sigma},
    \label{eq:Z-Score}
\end{equation}
where $x,x^{\prime}$ are the original and preprocessed input features, $\mu$ is the mean of the data, and $\sigma$ is the standard deviation of the data. The standardization procedure aims to transform the data into a distribution with a mean of 0 and a variance of 1. Another data preprocessing step involved is normalization, which is defined as
\begin{equation}
    x^{\prime}=\frac{x-\min (x)}{\max (x)},
    \label{eq:Normalize}
\end{equation}
where $x,x^{\prime}$ are the input features before and after normalization, $\min (x)$ is the minimum value of the input, and $\max (x)$ is the maximum of the input. The normalization process aims to transform the data into a distribution characterized by a minimum value of 0 and a maximum value of 1. The combination of these two preprocessing steps effectively reduces the data range, thereby accelerating the convergence of the neural network.

\subsubsection{Output feature}
\begin{figure}[h]
	\centering
	\includegraphics[width=0.5\linewidth]{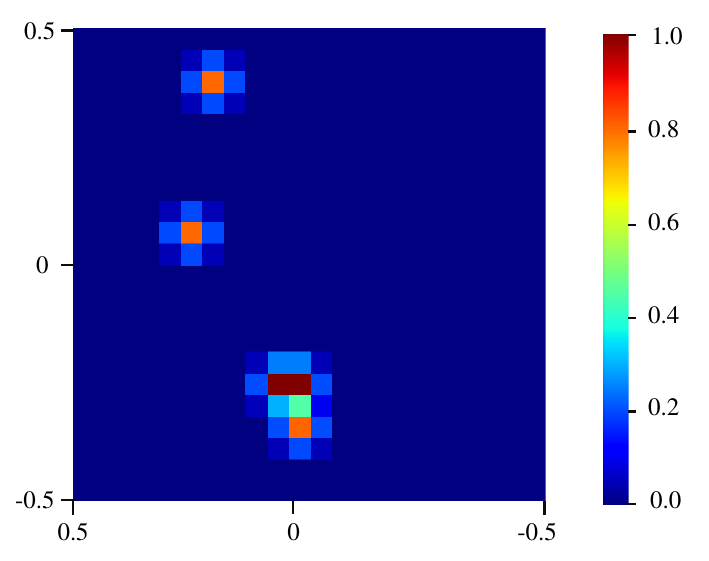}
	\caption{Output label of the CNN model.}
	\label{fig:generatedLabel}
\end{figure}
For an intuitive representation of the sound source position, we use the coordinates of subregions of the scanning grid as the output label, and the output value corresponding to each subregion is the probability of source existence. The output values are shown as the color of small subregions represented by square pixels which form a heat map of sources as in Figure~\ref{fig:generatedLabel}. When creating the training data, true source locations are marked on the individual subregions. However, since the strength of a real source may impact surrounding area, Gaussian blur function \cite{NIXON202083} is implemented to make the heat map of source distribution smooth.

\subsubsection{Loss function}

The loss function used in this study consists of two parts, as shown below
\begin{equation}
    F_{loss} = \frac{1}{K} \sum_{i=1}^{K} (q_{grid}^{(i)} - \hat{q}_{grid}^{(i)})^2 + \frac{1}{N} \sum_{j=1}^{N} (q_{source}^{(j)} - \hat{q}_{source}^{(j)})^2.
    \label{eq:sourceLocationloss}
\end{equation}
In this equation, $F_{loss}$ is the total prediction loss, $K$ is the number of the entire scanning grid points, and $N$ is the number of sound sources. $q_{grid}^{(i)}$ and $q_{source}^{(j)}$ are the true heat map values at the scanning grid points and at the real sound source position, respectively. Similarly, $\hat{q}_{grid}^{(i)}$ and $\hat{q}_{source}^{(j)}$ are the predicted heat map values at the scanning grid points and the real sound source position. It is important to note that while the first part of the mean square error loss (MSELoss) function penalizes the global prediction error, an emphasis on the penalization of the discrepancy between the true heat map value and the predicted value at source locations will lead to better prediction accuracy, hence, a source location loss is added as the second part of the total prediction loss. Test results will be given in Sec.~\ref{Sec:Results} to justify this customized loss function formulation.
\subsection{Accuracy evaluation algorithm}
\begin{algorithm}[!h]
	\caption{Get sound source coordiantes}
	\label{algo:getSourceCoordinates}
	\KwData{label $y$}
	\KwResult{coordinates}
	tolerance $t$, maximum value $m$\;
	$t\leftarrow 0.2$\;
	$m\leftarrow max(y)$\;
	\For{$y_i$ in $y$}{
		$value \leftarrow\lvert y_i - m \rvert$\;
		$indices \leftarrow findIndices(y_i)$\;
		\If{$value < t \times m$}
		{append indices to coordinates\;}
	}
	\Return{$coordinates$}\;
\end{algorithm}
As illustrated in Figure~\ref{fig:generatedLabel}, values corresponding to the sound sources are markedly higher and this distinctive feature can be leveraged in the sound source position extraction algorithm. We consider taking the maximum of the output and the values whose difference with the maximum are within 20\% as potential output of sound sources, and then extract their coordinates. This method, as shown in Algorithm \ref{algo:getSourceCoordinates}, can effectively identify the sound source position and neglect smear of the main lobe or unwanted sidelobe values. 

\begin{algorithm}[!hb]
	\caption{Get accuracy}
	\label{algo:getAccuracy}
	\KwData{true label $y$, predicted label $\hat{y}$}
	\KwResult{accuracy}
	true coordinates $tc$, predicted coordinates $pc$\;
	$tc\leftarrow getSourceCoordinates(y)$\;
	$pc\leftarrow getSourceCoordinates(\hat{y})$\;
	$intersection \leftarrow tc \cap pc$\;
	\If{$length_{tc} \neq 0$}
	{$accuracy \leftarrow length_{intersection} \div length_{tc}$\;}
	\Else{$accuracy \leftarrow 0$\;}
	\Return{$accuracy$}\;
\end{algorithm}
The ideal outcome for CNN-based SSL is that the number of predicted sources as well as their coordinates align with the truth. However, as shown by Lee et al. \cite{lee2022deep}, it often appears that a number of true sources are not captured by the prediction, and some predicted results are simply not aligned with the true sources. In their paper, "recall" and "precision" are defined to account for the percentage of correctly predicted sources with respect to the number of true sources and the number of predicted sources, respectively. There are two sets of quantities involved in assessing the accuracy, predicted and true source coordinates. The trusted results are from the intersection of the two sets, where part of the prediction aligns with the truth. In this paper we use the percentage of the number of trusted results within the total number of true sources to get prediction accuracy, as given in Algorithm~\ref{algo:getAccuracy}.

\section{Results}\label{Sec:Results}
In practice, the distance, frequency, and the number of sound sources are often unknown beforehand, and the signals received by the microphone array are usually noisy. One challenge is to train the CNN model such that it can adapt to these physical variations. Existing CNN-based SSL methods attempt to predict sound sources by fixing at least one of the above parameters and it is more straightforward to train a different model for each individual variable. For example, Xu et al. \cite{xu2021acoustic} fixed the microphone array-to-scanning grid distance and train different models for each frequency of interest. Improved upon this, the current CNN model is trained to adapt under different physical parameters. The adaptability and robustness are tested in respective cases. A test data set is generated for each testing case, and each data set consists of 4000 randomly generated sound field data. 

\subsection{Adaptability of the neural network}
\begin{figure}[!h]
	\centering
	\includegraphics[width=1.0\linewidth]{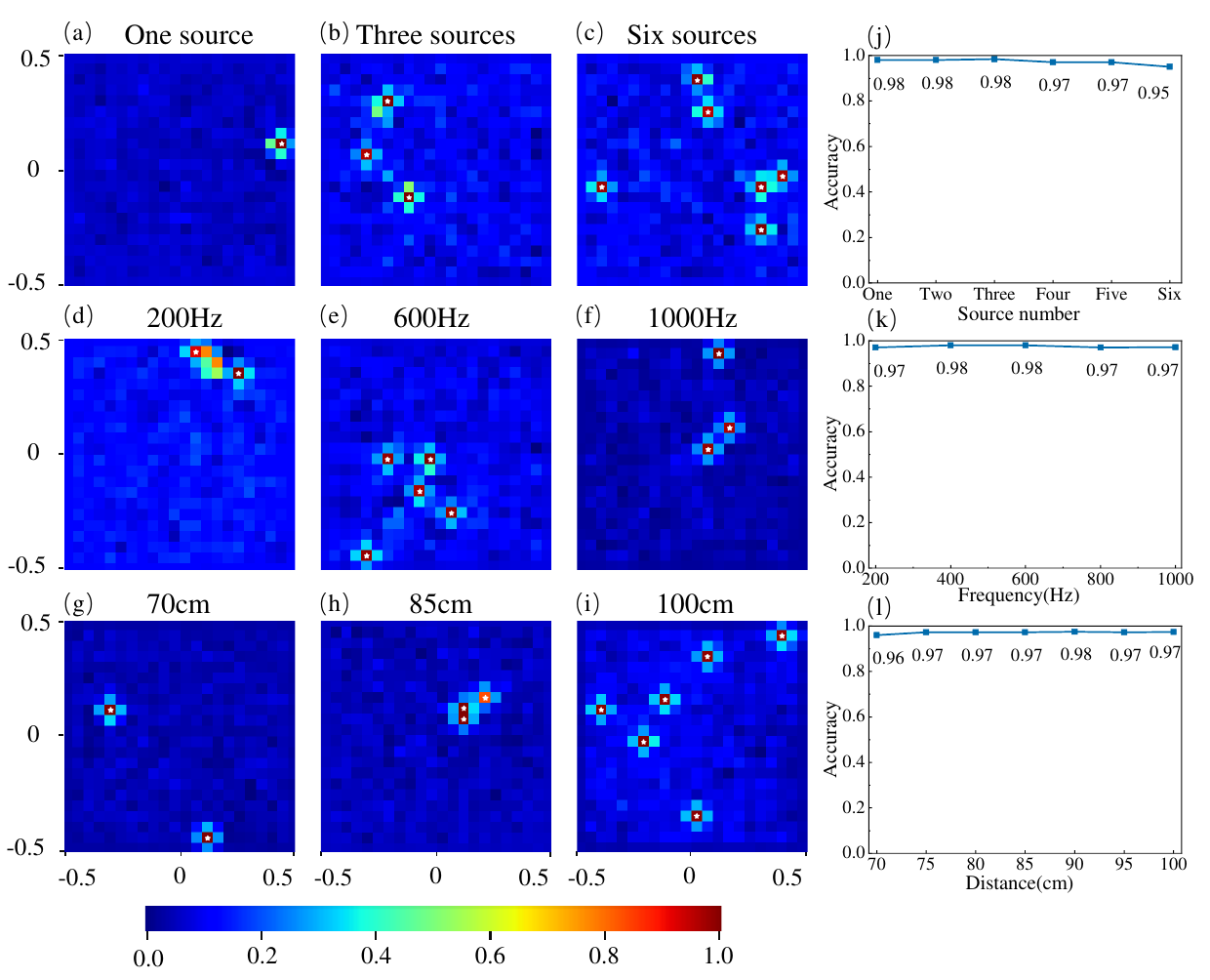}
	\caption{Adaptability of the CNN model across different scenarios.}
	\label{fig:resultsOfPlane30Width}
\end{figure}
To evaluate the adaptability of the neural network model, we first conduct tests with different numbers of sound sources. For a test data set with a given number of sound sources, each input contains the SPL values from a simulation with random frequency ranging from 200Hz to 1000Hz and microphone array-to-scanning grid distance ranging from 70cm to 100cm. We have tested up to six sound sources. The results of one, three, and six sound sources are presented in Figure~\ref{fig:resultsOfPlane30Width}(a-c) as heat maps, where the predicted sound sources are marked by pixels with magnitude close to one and the location of true sources are marked by small white stars. It is shown that sources are successfully identified in all three cases with much smaller off-peak values. Figure~\ref{fig:resultsOfPlane30Width}(j) shows that the model achieves an average localization accuracy exceeding 97\% across varying numbers of sound sources.

Next, we examine the adaptability of the model under different frequencies. We generate test datasets for a given frequency by randomly varying the number of sources and microphone array-to-scanning grid distances. Tests have been done from 200Hz to 1000Hz, and results randomly chosen from 200Hz, 600Hz, and 1000Hz cases are shown in Figure~\ref{fig:resultsOfPlane30Width} (d-f). At 200Hz, shown in Figure~\ref{fig:resultsOfPlane30Width} (d), the model accurately captures the sound source locations. However, it also gives some nearby off-peak values which would be filtered out by the set tolerance in Algorithm~\ref{algo:getSourceCoordinates} when assessing the accuracy. At 600 and 1000Hz, the model accurately predicts the positions of all sound sources, as depicted in Figure~\ref{fig:resultsOfPlane30Width}(e-f), without any occurrence of false sound source interference. Furthermore, the accuracy of the model's predictions at various frequencies, illustrated in Figure~\ref{fig:resultsOfPlane30Width} (k), demonstrates a consistent average accuracy exceeding 97\%.

To evaluate the neural network model's capacity under different microphone array-to-scanning grid distance, each data set with a given distance is generated with a random frequency and source number. Tests are done with the distance ranging from 70cm to 100cm, and results at 70cm, 85cm and 100cm are randomly chosen and shown in Figure~\ref{fig:resultsOfPlane30Width} (g-i). The outcomes reveal the model's adeptness in accurately capturing sound source locations across varying distances with negligible off-peak values. Figure~\ref{fig:resultsOfPlane30Width}(l) shows the model accuracy on varying distances across datasets generated for this test. Within the current configuration, the accuracy slightly increase with a farther distance, while maintaining an average accuracy around 97\%.

\subsection{Robustness of the neural network}
\begin{figure}[h]
	\centering
	\includegraphics[width=1.0\linewidth]{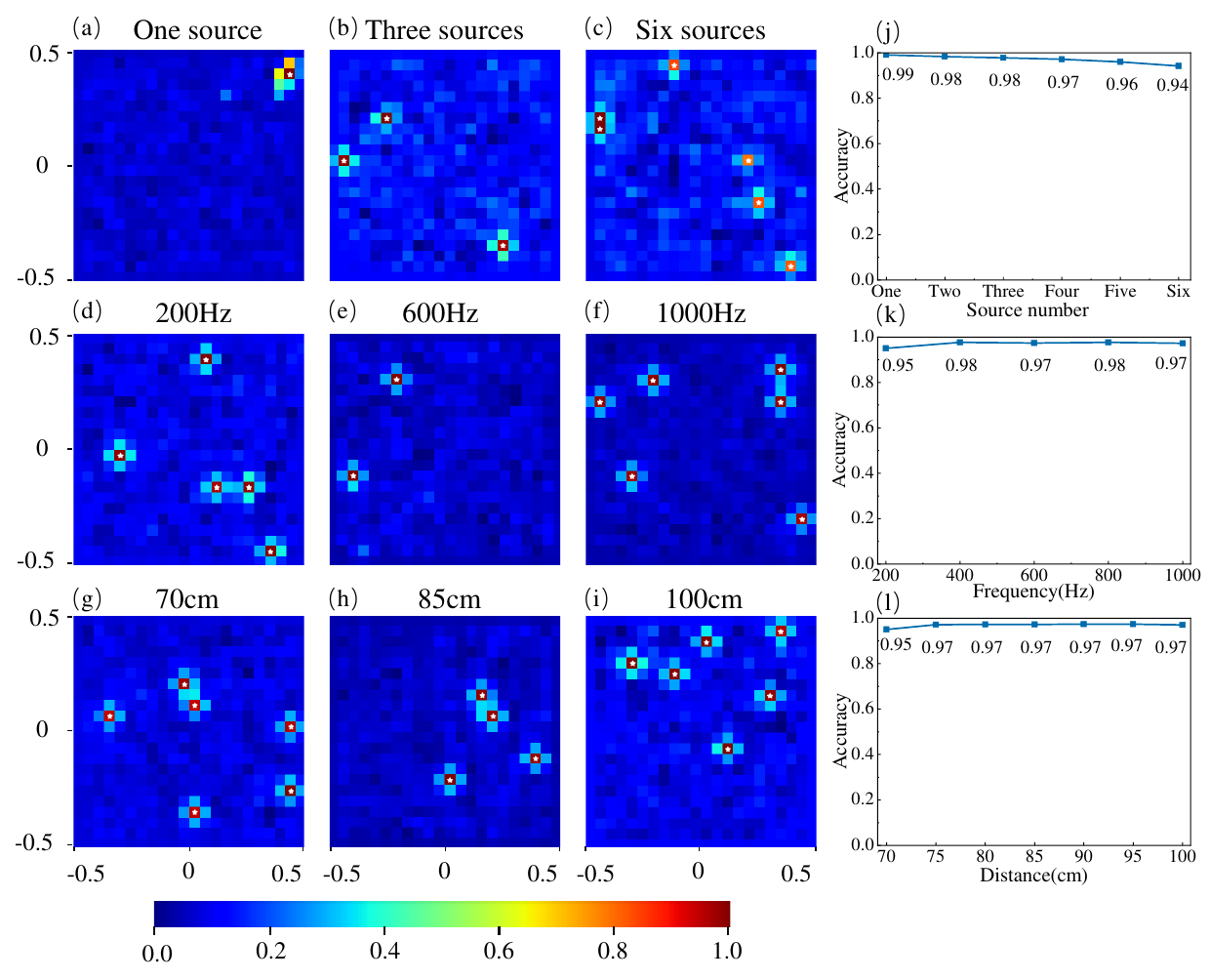}
	\caption{Robustness of the CNN model in the presence of noise.}
	\label{fig:addNoise}
\end{figure}
In this study, we seek to evaluate the robustness of neural network model under background Gaussian noise \cite{kumain2018efficient}. Noise is added as an input disturbance to each microphone in a data set, and it is similar to adding noise to an image, therefore we choose a SNR formulation commonly used in image processing \cite{Burger2022}, which is given as
\begin{equation}
    SNR = 10log_{10}(\frac{\sigma_{signal}^2}{\sigma_{noise}^2}),
    \label{eq:SNR}
\end{equation}
where $\sigma_{signal}$ and $\sigma_{noise}$ represent the standard deviation of the signal and the noise, respectively. Due to zero mean values in the Z-score standardized signal and Gaussian noise, the conventional SNR definition in terms of power ratio can be reduced into the above form.  In this study, SNR is set to $30dB$.

Robustness is tested under varying sound source numbers, frequencies, and microphone array-to-scanning grid distances. The results are presented in Figure~\ref{fig:addNoise}. Figures~\ref{fig:addNoise} (a-c) show that sound sources are accurately captured with close to one peak values marked in the heat maps, although with slight smearing in one source case. As shown in Figure~\ref{fig:addNoise} (j), the accuracy slightly drops when there are more sound sources, however, the average accuracy for different sound source numbers remains above 97\%.  

The results in Figures~\ref{fig:addNoise} (d-f) shows excellent source localization under different frequencies. However, Figure~\ref{fig:addNoise}(k) indicates a slight reduction in accuracy at 200Hz to 95\% and otherwise, the accuracy remains above 97\%. Figures~\ref{fig:addNoise} (g-i) consistently shows accurate source localization results at different microphone array-to-scanning grid distances in the presence of noise. Figure~\ref{fig:addNoise}(l) indicates marginal variation in accuracy above 97\% within the 75cm-100cm range, and a slight decline in accuracy at 70cm. 

\subsection{Performance at different accuracy values}

\begin{figure}[h]
    \centering
    \includegraphics[width=1.0\linewidth]{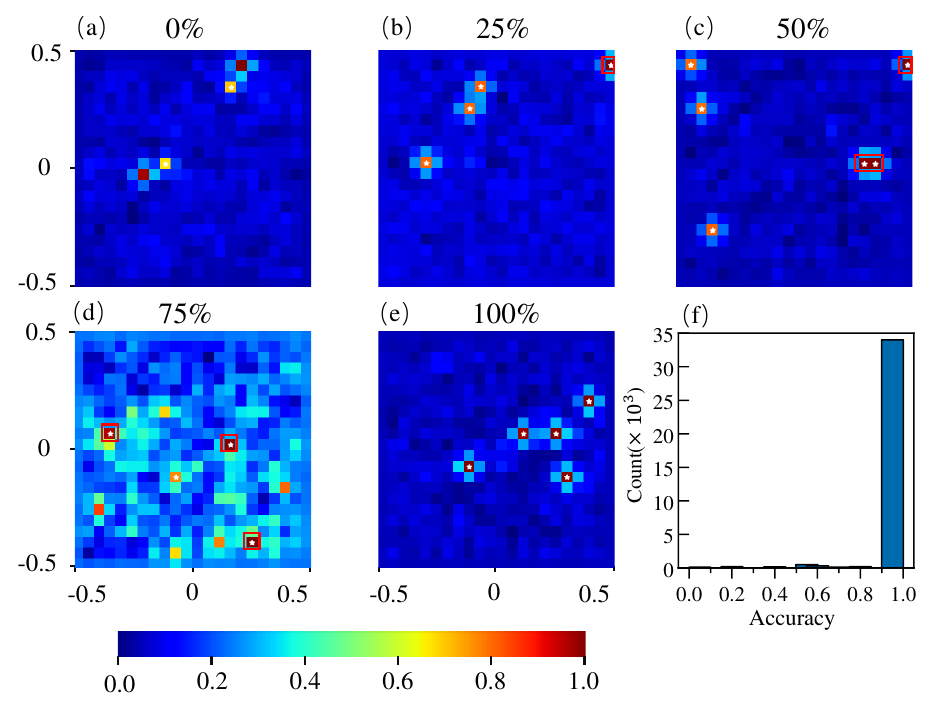}
    \caption{Visualization and statistics of SSL at different accuracy.}
    \label{fig:visualizationOfDifferentAccuracyValue}
\end{figure}

Since the accuracy under different occasions may vary, it is of interest to examine the actual performance of SSL at different accuracy values. Output heat maps are examined for accuracy values at 0\%, 25\%, 50\%, 75\% and 100\%, as shown in Figures~\ref{fig:visualizationOfDifferentAccuracyValue}(a-e). The correctly predicted source locations are marked by red boxes in 25\%, 50\% and 75\% results for clarification. The examples shown here represent either a common situation for that accuracy value or a case particular to the way accuracy is estimated. 

At 0\% accuracy, none of the sound sources are predicted correctly, however, the predicted sound sources are in close proximity to the true sound source locations. When accuracy reaches 25\%, only the sound source in the top right corner is accurately localized, evident by its close to one value in the heat map. Although Algorithm~\ref{algo:getSourceCoordinates} fails to correctly identify the other three sources, their predicted heat map values are obviously larger than subregions without a source. At 50\% and 75\% accuracy, heat maps are able to suggest all the source locations, however, only the values that satisfy the tolerance in Algorithm~\ref{algo:getSourceCoordinates} are counted, and thus give lower accuracy values. Interestingly, Figure~\ref{fig:visualizationOfDifferentAccuracyValue}(d) shown multiple heat map peaks, which exceeds the number of true sources. This usually happens when the sound field is complicated with the presence of multiple sources and background noise at low frequency. Heat map suggests there could be interference which contributes to the presence of pseudo-sources. When the accuracy reaches 100\%, the model accurately predicts the locations of all sound sources. The histogram of accuracy distribution calculated against the entire test datasets is shown in Figure~\ref{fig:visualizationOfDifferentAccuracyValue} (f), which suggests that instances of low prediction accuracy constitute only a small portion of total cases. Further improvement could be done by introducing a more flexible heat map peak value identification algorithm.

\subsection{Spatially distributed sound sources}
\begin{figure}[h]
    \centering
    \includegraphics[width=0.9\linewidth]{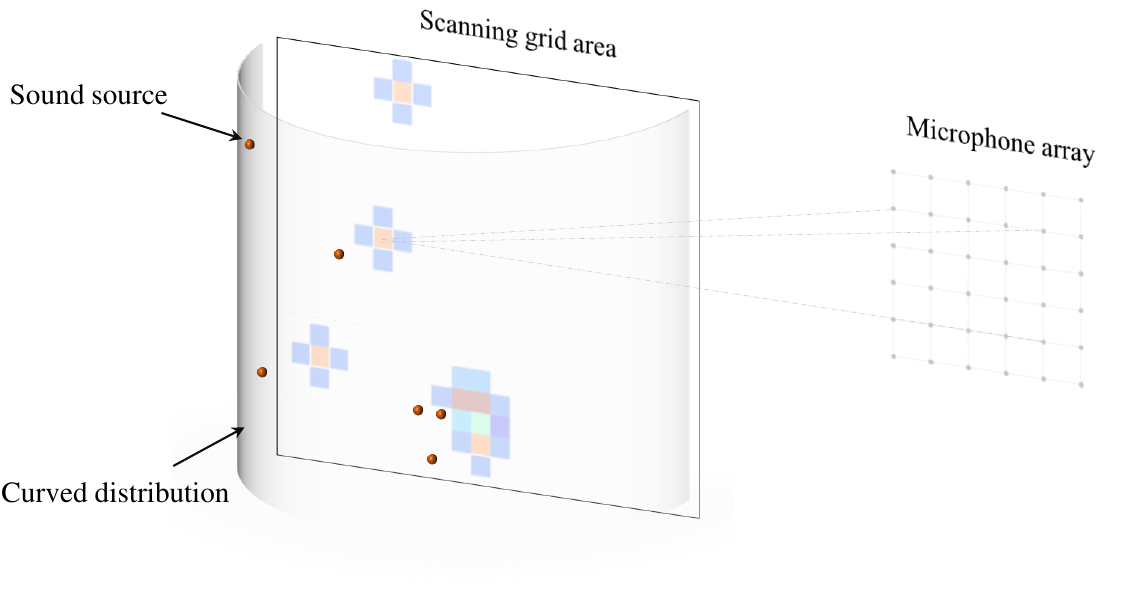}
    \caption{Schematic of the SSL problem with spatially distributed sound sources.}
    \label{fig:3dLocalizationStructure}
\end{figure}

\begin{figure}[htb]
	\centering
	\includegraphics[width=1.0\linewidth]{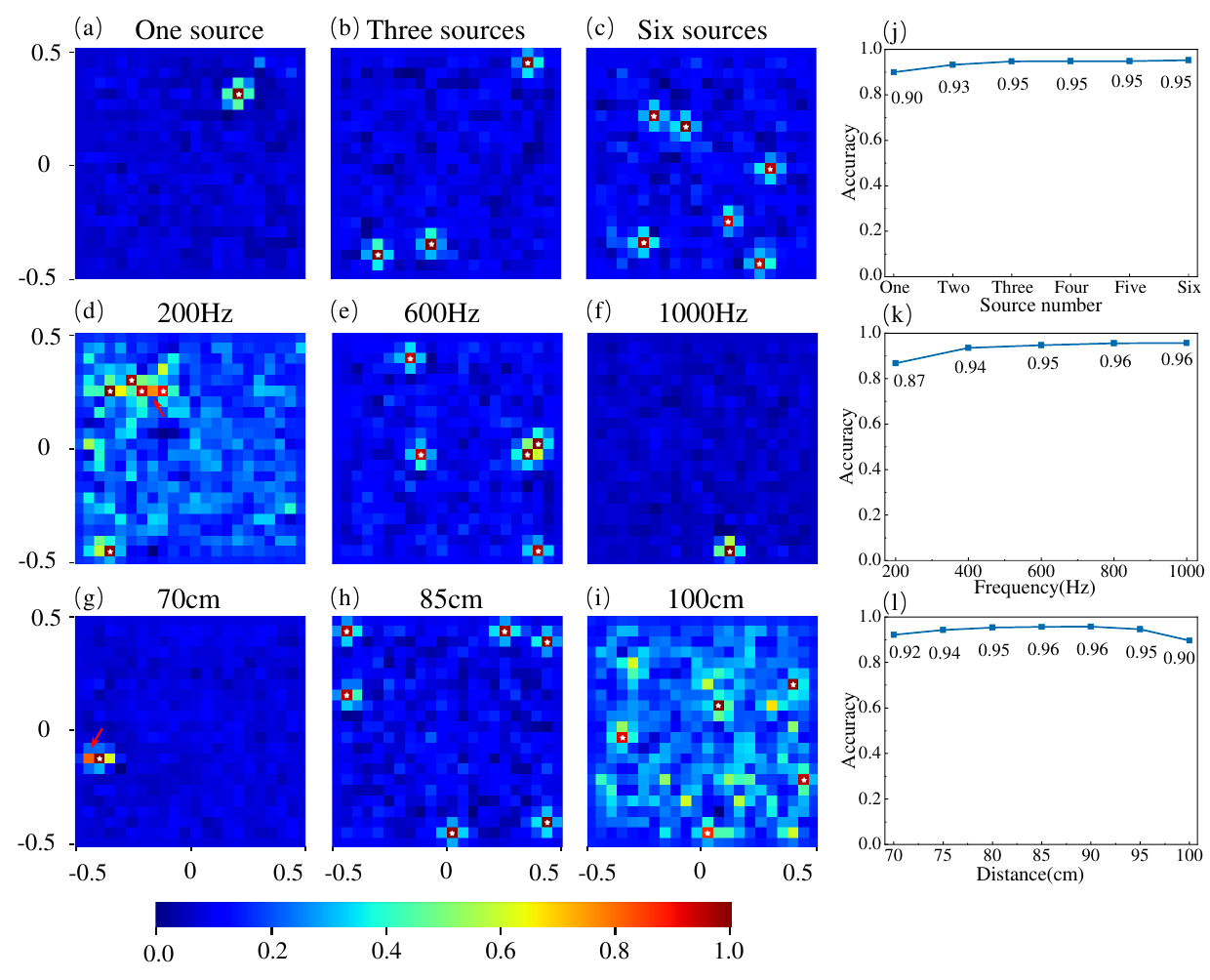}
	\caption{Adaptability of the CNN model for spatially distributed sound sources.}
	\label{fig:resultsOf3DLocalization}
\end{figure}

The neural network model's ability to adapt to sound fields generated by spatially distributed sources is evaluated. The sources are distributed along a cylindrical surface as shown in Figure~\ref{fig:3dLocalizationStructure}, therefore, the distance from the sources to the microphone array varies. However, the projection of source locations to the scanning grid is predicted by the neural network. The dimensions of the scanning grid and microphone array remain the same with those given in Figure~\ref{fig:planeSimulationStructure}. The results are shown in Figure~\ref{fig:resultsOf3DLocalization}.

We first look at the cases with different number of sources. As shown in  Figures~\ref{fig:resultsOf3DLocalization} (a-c), the model accurately predicts the positions of all sound sources.
From figure~\ref{fig:resultsOf3DLocalization} (j), we can see that at one source case, the accuracy of the model drops compared to the planar distribution case, with an accuracy of 90\%. Between two sources and five sources, the average accuracy is 94\%, slightly decreased compared with the planar case. At six sources, the accuracy of the model is almost the same as that of the planar distribution case, with an accuracy of 95\%. 

Figures~\ref{fig:resultsOf3DLocalization}(d-f) show the prediction of the sound source locations under various frequencies. Although at 200Hz the model successfully captures all real source locations, an additional prediction presents, as indicated by the red arrow in Figure~\ref{fig:resultsOf3DLocalization}(d). Apparently, in the current five sources case, sound field pattern could be complicated for the model to remove pseudo-sources such as the one shown in the heat map. At 600Hz and 1000Hz, shown in Figure~\ref{fig:resultsOf3DLocalization}(e-f), the model accurately predicts the locations of all sound sources. The accuracy of the model at different frequencies is shown in Figure~\ref{fig:resultsOf3DLocalization}(k), which shows that in 200Hz, the model has a drop in accuracy compared to the planar distribution case due to the increased complexity of source distribution in space, with an accuracy of 87\%. However, as the frequency increases, the accuracy of the model also increases, with an average accuracy of 94\% which is almost the same as that of the planar distribution case.

Figures~\ref{fig:resultsOf3DLocalization}(g-i) show the models adaptability to different microphone array-to-scanning grid distances with spatially distributed sources. It is seen that all sound sources locations are accurately predicted, although smear of the main lobe happens in 70cm case, as indicated by the red arrow in Figure~\ref{fig:resultsOf3DLocalization}(g). Figure~\ref{fig:resultsOf3DLocalization}(l) shows the highest accuracy within distances of 75-95cm, with a slightly decreased accuracy at closer (70cm) and farther (100cm) distances. The average accuracy stands at approximately 94\%, merely 3\% lower than that of the planar distribution case. In summary, the neural network model has demonstrated impressive adaptability to spatial source distribution, despite a slight decrease in accuracy compared to the planar distribution cases.

\subsection{Effectiveness of the customized loss function}

The effectiveness of the customized loss function, as illustrated in Equation~\ref{eq:sourceLocationloss}, is evaluated in this section. The configuration of the microphone array and scanning grid keeps the same as the previous case as depicted in Figure~\ref{fig:3dLocalizationStructure}. These results are compared with those model trained using only the MSELoss function. 
\begin{figure}[htbp]
    \centering
    \includegraphics[width=1.0\linewidth]{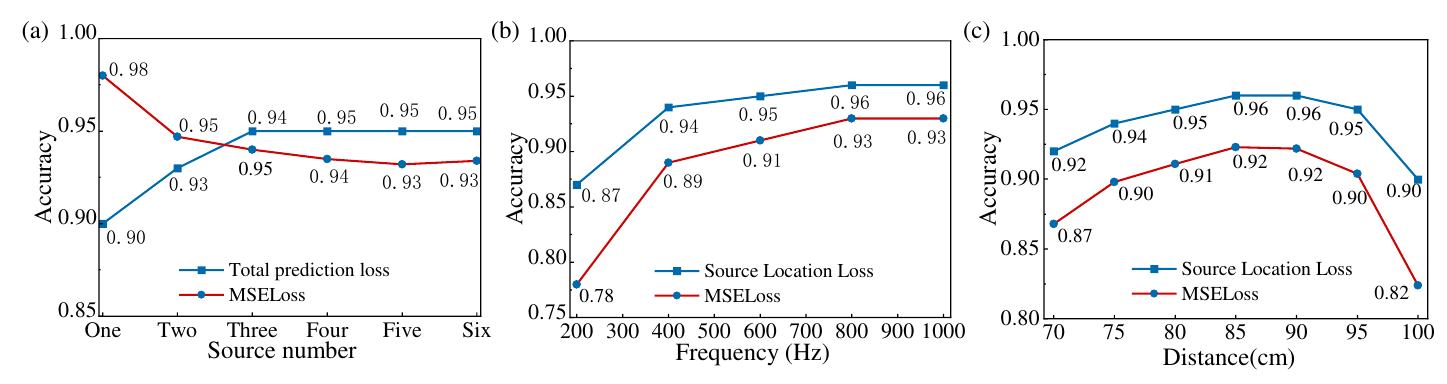}
    \caption{Effectiveness of the customized loss function.}
    \label{fig:effectivenessOfLossFunction}
\end{figure}

From Figure~\ref{fig:effectivenessOfLossFunction}(a), we can find that the accuracy of the model trained with MSELoss is higher than that of the model trained with the customized loss function when there is one or two sound sources. However, for three to six sources, the model trained with the customized loss function outperforms the model trained with MSELoss. Furthermore, Figures~\ref{fig:effectivenessOfLossFunction}(b, c) demonstrate that the model trained with the customized loss function consistently outperforms that trained with MSELoss across all frequencies and distances. These findings suggest that the customized loss function effectively enhances the accuracy of the neural network model in SSL. 

\subsection{Comparison with acoustic beamforming algorithms}

\begin{figure}[htbp]
    \centering
    \includegraphics[width=1.0\linewidth]{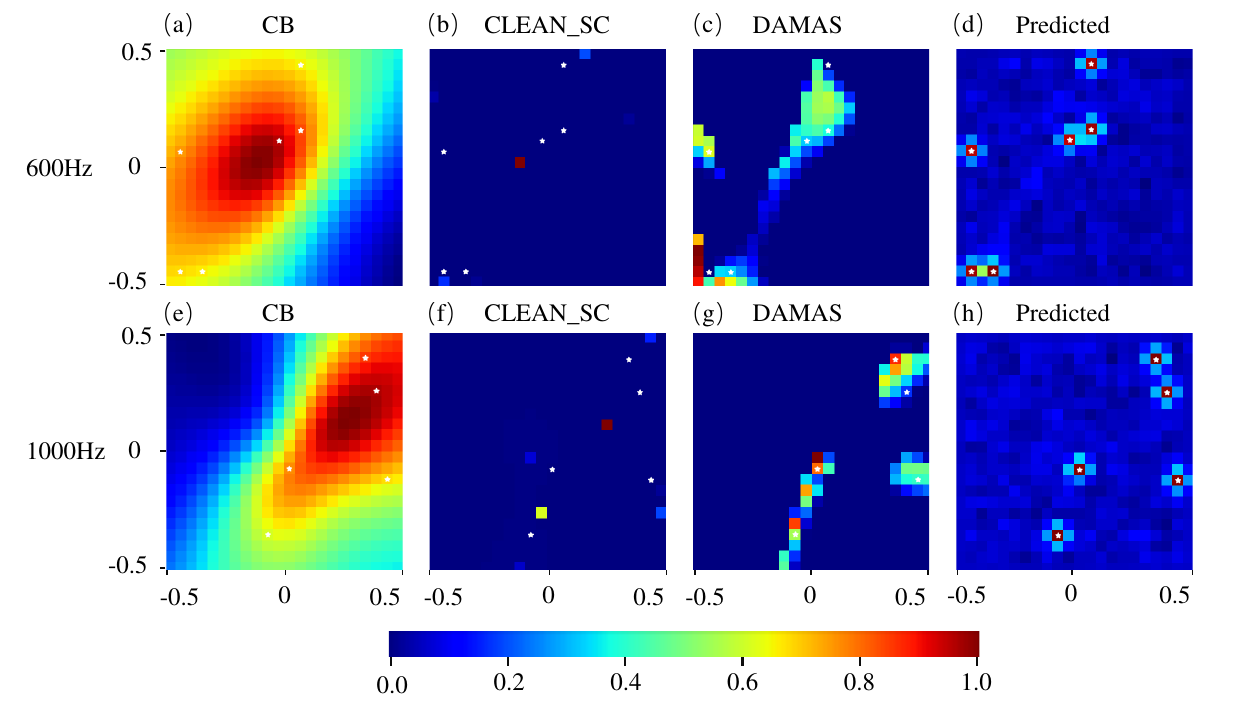}
    \caption{Comparison of the CNN model with acoustic beamforming algorithms.}
    \label{fig:comparasionWithTraditionalAlgorithms}
\end{figure}

It is a common practice to employ acoustic beamforming algorithms as a benchmark for deep learning models. In this study acoustic beamforming algorithms are implemented with the same simulation conditions to the deep learning model, as illustrated in Figure~\ref{fig:planeSimulationStructure}, where the microphone array-to-scanning grid distance is set to 1$m$. It is shown in Figures~\ref{fig:comparasionWithTraditionalAlgorithms}(a, e) that the CB algorithm is able to approximate sound source distribution at both 600Hz and 1000Hz but not able to ascertain the sound source locations. CLEAN-SC on the other hand gives a very rough estimation of the sound source locations in the proximity of true locations, and the heat maps suggest a better approximation at 1000Hz as shown in Figures~\ref{fig:comparasionWithTraditionalAlgorithms}(b, f). DAMAS algorithm, as shown in Figures~\ref{fig:comparasionWithTraditionalAlgorithms}(c, g), does not accurately determine the locations of all sources, except the two at 1000Hz, but it does establish the overall distribution of sources via heat map values. Our neural network model results in Figure~\ref{fig:comparasionWithTraditionalAlgorithms}(d, h) precisely predict the locations of all sound sources at both 600Hz and 1000Hz, showing significantly enhanced accuracy at low frequencies compared to acoustic beamforming algorithms.

In summary, acoustic beamforming algorithms, such as DAMAS and CLEAN-SC, are capable of roughly identifying the potential locations of a sound source at frequencies of 600Hz and 1000Hz with higher accuracy than CB. In comparison, our neural network model possesses the ability to effortlessly ascertain both the position and number of sound sources.

\section{Conclusion}

This paper introduces a CNN model with a customized loss function capable of achieving highly accurate SSL with a small-sized microphone array. The adaptability to various numbers and frequencies of sound sources, as well as different distances is tested. Furthermore, the model's robustness in SSL under specific SNR and its accuracy in locating sources in more complex scenes are examined. It is shown that for all the planar distributed cases test accuracy can reach an average of 97\%. For the cases where sources are distributed on a cylindrical surface, an average around 94\% accuracy can be obtained. For the cases with lower prediction accuracy, although exact source locations are not captured, peak values in heat map emerge in the close proximity of true locations, or appear with slightly lower than accepted peak values defined in Algorithm \ref{algo:getSourceCoordinates} at correct locations. The neural network structure and associated parameters presented in this study demonstrate the capacity for adaptive, high-resolution SSL at low frequencies, surpassing existing algorithms. Moreover, the lightweight structure of the neural network and the characteristics of its input and output features contribute to its fast inference speed, thereby advancing the potential for real-time SSL.

\section{Acknowledgement}

The authors would like to thank the financial support from National Natural Science Foundation of China (Project Nos. 12372198 and 12302111).

\section*{CRediT author statement}
\textbf{Wenbo Ma:} Methodology, Investigation, Software, Data Curation, Writing - Original Draft, Visualization. \textbf{Yan Lu:} Conceptualization, Methodology, Investigation, Writing - Review \& Editing, Supervision, Funding acquistion. \textbf{Yijun Liu:} Conceptualization, Resources, Supervision, Project administration, Funding acquisition.

\bibliographystyle{elsarticle-num}

\begin{thebibliography}{10}
	\expandafter\ifx\csname url\endcsname\relax
	\def\url#1{\texttt{#1}}\fi
	\expandafter\ifx\csname urlprefix\endcsname\relax\def\urlprefix{URL }\fi
	\expandafter\ifx\csname href\endcsname\relax
	\def\href#1#2{#2} \def\path#1{#1}\fi
	
	\bibitem{lee2012case}
	G.-S. Lee, C.~Cheong, S.-H. Shin, S.-S. Jung, A case study of localization and
	identification of noise sources from a pitch and a stall regulated wind
	turbine, Applied Acoustics 73~(8) (2012) 817--827.
	
	\bibitem{ballesteros2015noise}
	J.~A. Ballesteros, E.~Sarradj, M.~D. Fernandez, T.~Geyer, M.~J. Ballesteros,
	Noise source identification with beamforming in the pass-by of a car, Applied
	Acoustics 93 (2015) 106--119.
	
	\bibitem{cabada2015acoustic}
	E.~C. Cabada, N.~Hamzaoui, Q.~Leclere, J.~Antoni, Acoustic imaging applied to
	fault detection in rotating machine, in: Surveillance 8, 2015.
	
	\bibitem{chazan2019multi}
	S.~E. Chazan, H.~Hammer, G.~Hazan, J.~Goldberger, S.~Gannot, Multi-microphone
	speaker separation based on deep doa estimation, in: 2019 27th European
	Signal Processing Conference (EUSIPCO), IEEE, 2019, pp. 1--5.
	
	\bibitem{songgong2020robust}
	K.~SongGong, H.~Chen, Robust indoor speaker localization in the circular
	harmonic domain, IEEE Transactions on Industrial Electronics 68~(4) (2020)
	3413--3422.
	
	\bibitem{liu2018sound}
	G.~Liu, S.~Yuan, J.~Wu, R.~Zhang, A sound source localization method based on
	microphone array for mobile robot, in: 2018 Chinese Automation Congress
	(CAC), IEEE, 2018, pp. 1621--1625.
	
	\bibitem{li2016reverberant}
	X.~Li, L.~Girin, F.~Badeig, R.~Horaud, Reverberant sound localization with a
	robot head based on direct-path relative transfer function, in: 2016 IEEE/RSJ
	International Conference on Intelligent Robots and Systems (IROS), IEEE,
	2016, pp. 2819--2826.
	
	\bibitem{merino2019review}
	R.~Merino-Mart{\'\i}nez, P.~Sijtsma, M.~Snellen, T.~Ahlefeldt, J.~Antoni, C.~J.
	Bahr, D.~Blacodon, D.~Ernst, A.~Finez, S.~Funke, et~al., A review of acoustic
	imaging methods using phased microphone arrays: Part of the “aircraft noise
	generation and assessment” special issue, CEAS Aeronautical Journal 10
	(2019) 197--230.
	
	\bibitem{chiariotti2019acoustic}
	P.~Chiariotti, M.~Martarelli, P.~Castellini, Acoustic beamforming for noise
	source localization--reviews, methodology and applications, Mechanical
	Systems and Signal Processing 120 (2019) 422--448.
	
	\bibitem{yardibi2010uncertainty}
	T.~Yardibi, C.~Bahr, N.~Zawodny, F.~Liu, L.~Cattafesta~III, J.~Li, Uncertainty
	analysis of the standard delay-and-sum beamformer and array calibration,
	Journal of Sound and Vibration 329~(13) (2010) 2654--2682.
	
	\bibitem{rayleigh1879xxxi}
	Rayleigh, Xxxi. investigations in optics, with special reference to the
	spectroscope, The London, Edinburgh, and Dublin Philosophical Magazine and
	Journal of Science 8~(49) (1879) 261--274.
	
	\bibitem{herold2015approach}
	G.~Herold, E.~Sarradj, An approach to estimate the reliability of microphone
	array methods, in: 21st AIAA/CEAS aeroacoustics conference, 2015, p. 2977.
	
	\bibitem{brooks2006deconvolution}
	T.~F. Brooks, W.~M. Humphreys, A deconvolution approach for the mapping of
	acoustic sources (damas) determined from phased microphone arrays, Journal of
	sound and vibration 294~(4-5) (2006) 856--879.
	
	\bibitem{sijtsma2007clean}
	P.~Sijtsma, Clean based on spatial source coherence, International journal of
	aeroacoustics 6~(4) (2007) 357--374.
	
	\bibitem{castellini2013average}
	P.~Castellini, A.~Sassaroli, A.~Paonessa, A.~Peiffer, A.~Roeder, Average
	beamforming in reverberant fields: Application on helicopter and airplane
	cockpits, Applied acoustics 74~(1) (2013) 198--210.
	
	\bibitem{zohourian2017binaural}
	M.~Zohourian, G.~Enzner, R.~Martin, Binaural speaker localization integrated
	into an adaptive beamformer for hearing aids, IEEE/ACM Transactions on Audio,
	Speech, and Language Processing 26~(3) (2017) 515--528.
	
	\bibitem{krause2021feature}
	D.~Krause, A.~Politis, K.~Kowalczyk, Feature overview for joint modeling of
	sound event detection and localization using a microphone array, in: 2020
	28th European Signal Processing Conference (EUSIPCO), IEEE, 2021, pp. 31--35.
	
	\bibitem{vecchiotti2019end}
	P.~Vecchiotti, N.~Ma, S.~Squartini, G.~J. Brown, End-to-end binaural sound
	localisation from the raw waveform, in: ICASSP 2019-2019 IEEE International
	Conference on Acoustics, Speech and Signal Processing (ICASSP), IEEE, 2019,
	pp. 451--455.
	
	\bibitem{pujol2021beamlearning}
	H.~Pujol, E.~Bavu, A.~Garcia, Beamlearning: An end-to-end deep learning
	approach for the angular localization of sound sources using raw multichannel
	acoustic pressure data, The Journal of the Acoustical Society of America
	149~(6) (2021) 4248--4263.
	
	\bibitem{ma2019phased}
	W.~Ma, X.~Liu, Phased microphone array for sound source localization with deep
	learning, Aerospace Systems 2~(2) (2019) 71--81.
	
	\bibitem{shimada2021accdoa}
	K.~Shimada, Y.~Koyama, N.~Takahashi, S.~Takahashi, Y.~Mitsufuji, Accdoa:
	Activity-coupled cartesian direction of arrival representation for sound
	event localization and detection, in: ICASSP 2021-2021 IEEE International
	Conference on Acoustics, Speech and Signal Processing (ICASSP), IEEE, 2021,
	pp. 915--919.
	
	\bibitem{lee2022deep}
	S.~Y. Lee, J.~Chang, S.~Lee, Deep learning-enabled high-resolution and fast
	sound source localization in spherical microphone array system, IEEE
	Transactions on Instrumentation and Measurement 71 (2022) 1--12.
	
	\bibitem{grumiaux2022survey}
	P.-A. Grumiaux, S.~Kiti{\'c}, L.~Girin, A.~Gu{\'e}rin, A survey of sound source
	localization with deep learning methods, The Journal of the Acoustical
	Society of America 152~(1) (2022) 107--151.
	
	\bibitem{niu2019deep}
	H.~Niu, Z.~Gong, E.~Ozanich, P.~Gerstoft, H.~Wang, Z.~Li, Deep-learning source
	localization using multi-frequency magnitude-only data, The Journal of the
	Acoustical Society of America 146~(1) (2019) 211--222.
	
	\bibitem{xu2021acoustic}
	P.~Xu, E.~J. Arcondoulis, Y.~Liu, Acoustic source imaging using densely
	connected convolutional networks, Mechanical Systems and Signal Processing
	151 (2021) 107370.
	
	\bibitem{bengio2017deep}
	Y.~Bengio, I.~Goodfellow, A.~Courville, Deep learning, Vol.~1, MIT press
	Cambridge, MA, USA, 2017.
	
	\bibitem{lin2021acoustic}
	H.~Lin, T.~Bengisu, Z.~P. Mourelatos, H.~Lin, T.~Bengisu, Z.~P. Mourelatos,
	Acoustic waves from spherical sources, Lecture Notes on Acoustics and Noise
	Control (2021) 137--165.
	
	\bibitem{he2015delving}
	K.~He, X.~Zhang, S.~Ren, J.~Sun, Delving deep into rectifiers: Surpassing
	human-level performance on imagenet classification, in: Proceedings of the
	IEEE international conference on computer vision, 2015, pp. 1026--1034.
	
	\bibitem{ioffe2015batch}
	S.~Ioffe, C.~Szegedy, Batch normalization: Accelerating deep network training
	by reducing internal covariate shift, in: International conference on machine
	learning, pmlr, 2015, pp. 448--456.
	
	\bibitem{kingma2014adam}
	D.~P. Kingma, J.~Ba, Adam: A method for stochastic optimization, arXiv preprint
	arXiv:1412.6980 (2014).
	
	\bibitem{NIXON202083}
	M.~S. Nixon, A.~S. Aguado, 3 - image processing, in: M.~S. Nixon, A.~S. Aguado
	(Eds.), Feature Extraction and Image Processing for Computer Vision (Fourth
	Edition), fourth edition Edition, Academic Press, 2020, pp. 83--139.
	
	\bibitem{kumain2018efficient}
	S.~C. Kumain, M.~Singh, N.~Singh, K.~Kumar, An efficient gaussian noise
	reduction technique for noisy images using optimized filter approach, in:
	2018 first international conference on secure cyber computing and
	communication (ICSCCC), IEEE, 2018, pp. 243--248.
	
	\bibitem{Burger2022}
	W.~Burger, M.~J. Burge, Color Images, Springer International Publishing, Cham,
	2022, pp. 375--423.
	
\end{thebibliography}

\end{document}